\documentclass{article}

\usepackage{graphicx,amsmath,bm}

\begin{document}

\begin{center}

{\boldmath \Large \bf Determining the structure of $Hqt$ couplings \\
at the LHC}

\bigskip
\bigskip


{\large
Saurabh D. Rindani\footnote{Email: sdrindani@gmail.com}
}

\bigskip

{\large \it Theoretical Physics Division, Physical
Research Laboratory\\ Navrangpura, Ahmedabad 380009, India
}
\end{center}
\nopagebreak

\bigskip\bigskip


\begin{center}

{\bf Abstract}

\end{center}

\bigskip

Experiments at the Large Hadron Collider (LHC) have found that
 the Higgs boson discovered in 2012 has properties largely 
consistent with
predictions from the standard model (SM). However, the
determination of the couplings of the Higgs boson to various other SM
particles is not very precise, and it is possible that 
future experiments with enhanced accuracy may be able to detect
deviations from the SM if any. 
One category of couplings studied at the LHC is Higgs
flavour-changing couplings to a top-quark and a light-quark ($u$ or $c$)
combination, 
which vanishes at tree level
in the SM, but could be induced by interactions beyond the SM.
Experiments have put a limit on the magnitude of these couplings, only 
one coupling assumed nonzero at a time. 
We investigate here to what
extent the relative magnitudes of the $Hut$ and $Hct$ couplings and
their chirality structure can be determined by means of more detailed
kinematics at a future higher-luminosity version of the LHC
in the process of $tH$ production.
In
particular, we find that the top-quark polar-angle distribution in $tH$
production could reveal the relative magnitudes of the $Hut$ and $Hct$
couplings, while the azimuthal distribution of the charged leptons 
arising from top decay can be used for a determination of the chiral
structure of the $Hut$ coupling.


\bigskip

Pacs: {12.10.Dm; 12.60.-i; 14.65.Ha; 14.80.Cp}

\bigskip

\bigskip

\newpage


\section{Introduction}

Experiments at the Large Hadron Collider (LHC) have found that
 the Higgs boson discovered in 2012 has properties largely 
consistent with
predictions from the standard model (SM). However, the accuracy in the
determination of the couplings of the Higgs boson to various other SM
particles is still somewhat limited,  and it is possible that 
future experiments with enhanced accuracy may be able to detect
possible deviations from the SM. 

One of the studies at the LHC is of a possible Higgs
flavour-changing coupling to a top-quark and light-quark combination, 
which vanishes at tree level
in the SM, but could be induced by interactions beyond the SM. 
Such interactions are found, for example, in models with 
warped extra dimensions \cite{Azatov:2009na},
compositeness models \cite{Azatov:2014lha},
two-Higgs doublet models (2HDM)
\cite{Aguilar-Saavedra:2002phh, Branco:2011iw, Bejar:2000ub,
Guasch:1999jp},
supersymmetric models with R-parity
violation \cite{Cao:2007dk}, and quark-singlet models 
\cite{Cao:2014udj}.
The $t\to cH$ 
coupling in particular can be enhanced in 2HDM
models  \cite{Eilam:2001dh}, including models with flavor-violating
Yukawa couplings \cite{Chen:2013qta} or 2HDM for the top quark
\cite{Baum:2008qm, Kao:2011aa, Altunkaynak:2015twa}.

Experiments have put a limit on the magnitude of such couplings. The
processes studied in these  
experiments are $tH$ associated production, $t\bar t$ pair production
and $t$ decay into $H$ and a lighter quark \cite{
ATLAStau, ATLAS:2018jqi, ATLAS:2022gzn, CMSJHEP, CMSPRL} at the
centre-of-mass (cm) energy of 13 TeV. The latest analysis in 
\cite{ATLAS:2022gzn, CMSPRL} is for integrated luminosity of
137 fb$^{-1}$. 
The light quarks contributing in the flavour-changing couplings are the
$u$ quark and the $c$ quark.

In these
experimental determinations, it has been assumed that one particular
coupling contributes, and it has not been possible to distinguish
between a coupling involving a $u$ quark and one involving a $c$ quark.
The other feature  of an effective $tH$ coupling to a light
quark which is not determined is the relative contribution of 
left-handed and right-handed
chiralities. Since the measured cross sections and top decay branching
ratios depend only on the sum of the squares of the left-handed and
right-handed couplings, the experiments either put a limit on this
combination of couplings, or on a coupling assumed of 
to be of definite chirality (right-handed).

It is hoped that future higher-luminosity experiment
will be able to shed more light on the relative magnitudes of the 
$u$-quark and $c$-quark couplings, as well as relative contributions of
left-handed and right-handed couplings.

We investigate here to what
extent the relative magnitudes of the $Hut$ and $Hct$ couplings and
their chirality structure can be determined by means of more detailed
kinematics at a future higher-luminosity run of the LHC. 
Surely, a detailed fit to the kinematics, if permitted by the 
statistics, would certainly reveal details of couplings. However, we try
to look at a few kinematic features, rather than
fits, which would hopefully be possible to study with an integrated 
luminosity higher than what been has been possible so far, but within
what is anticipated for future runs of the LHC. 
In
particular, we study the top-quark polar-angle distribution in $tH$
production. We find that it could reveal the relative magnitudes of the 
$Hut$ and $Hct$
couplings.  We also investigate the polar and azimuthal distributions
 of the charged leptons 
arising from top decay. Our study shows that while the polar
distribution is insensitive to the chiral structure of the couplings, 
the azimuthal
distribution can be used for a determination of the chiral
structure of the $Hut$ coupling. The corresponding measurement for the
$Hct$ coupling  is insensitive to the chiral structure.

\section{\boldmath The process of $tH$ production through 
$Hqt$ couplings}

Before going to the details of our numerical analysis, we describe here
features of the $tH$ production process in the presence of
flavour-changing $Hqt$ couplings. 

Associated single-top Higgs production process, without additional jets,
 is not possible in the SM because of the absence of flavour-changing
neutral currents.  However, if flavour-changing Higgs interactions to
quarks occur in extensions of the SM, they can give rise to such a
final state through the partonic sub-processes $ug \to tH$ or 
$cg \to tH$.
Corresponding conjugate processes would give rise to the conjugate $\bar
tH$ final state. In LHC experiments, these processes are studied taking
care to avoid additional partons in the final state, so that the SM
contributions are not included.

The effective Lagrangian describing the fla\-vour-cha\-nging Higgs
interactions is given by
\begin{equation}\label{effLag}
{\cal L} = \sum_{q = u,c} \frac{g}{\sqrt{2}} \bar t \kappa_{Hqt}
		(F_L^{Hq}P_L + F_R^{Hq}P_R)q H + h.c.,
\end{equation}
where $g$ is the weak coupling constant, $P_L$ and $P_R$ are the usual
left and right chirality projection matrices, $\kappa_{Hqt}$ is the
effective overall coupling for the quark $q$,  and 
$F_L^{Hq}$ and $F_R^{Hq}$ are complex parameters satisfying
$|F_L^{Hq}|^2 + |F_R^{Hq}|^2 = 1$.

The contribution of each light quark to the total cross section is
independent of the chirality of the couplings, and is determined by
the value of $\kappa_{Hqt}$.
The cross section thus can put a limit on the relevant $\kappa_{Hqt} $
if only
$q=u$ or only $q=c$ contribution is present. 
However, in the case of both $u$ and $c$ contributions being present, 
the experimental limit on the cross section 
does not determine individual $u$ quark and $c$ quark couplings.
What the limit on the cross section can restrict is the combination
$\kappa_{Hut}^2 \sigma_u + \kappa_{Hct}^2 \sigma_c$, where 
$\sigma_{u,c}$ 
is the cross
section for the $u, c$ quark contribution. 
Experiments \cite{CMSJHEP,CMSPRL} determine the limit on the coupling  
assuming,
firstly, only right-handed coupling and secondly, either a $u$-quark
contribution or a $c$-quark contribution. 

We now present the numerical results for the polar distribution of the
top quark and the polar and azimuthal distributions of the charged
lepton from top-quark decay, and discuss how these can be used to
determine more details of the flavour-violating couplings.
The calculations are based on an analytical calculation at the partonic
level of the spin density
matrix for the production of a polarized top quark, and for its decay 
into a $b\ell^+\nu_{\ell}$ (and similarly for the top anti-quark) \cite{toppolazi}.
Integration over the parton densities and over relevant phase-space
variables is carried out numerically.

In view of the identical form of the effective $Hqt$ interactions 
chosen for the $u$ quark and the $c$ quark, it is natural to wonder what
gives rise to the difference in angular distributions of the top quark
and of the charged leptons arising from the $u$-quark and $c$-quark
contributions. One possibility is, of course, the difference in the
	masses of these quarks. However, in our numerical calculations,
the $u$ and $c$ masses have been neglected, so that possibility as the
cause of the differences in the angular distributions is ruled out. 
It is clear that with the $u$ and $c$ quarks assumed massless, 
there would be no difference between their contributions at the parton
level. 

The
other possible reason is the differences in the parton distribution
functions (PDF's) 
used for the $u$ and $c$ quarks. The differences in the PDF's are
certainly responsible for differences in the contributions to the total
cross section from these quarks, since the $u$ PDF having a valence
component would contribute a larger amount to the total cross
section. However, the differences in the $u$ and $c$ PDF's can also give
rise to differences in the shapes of angular distributions because the
$u$ and $\bar u$ PDF's would be significantly different in magnitude and
shape from each other, 
whereas the the $c$ and $\bar c$ PDF's would be similar to each
other in magnitude and shape. 

\section{Numerical Results}

Since the study we propose needs some  kinematic details, 
a future LHC run with a higher luminosity will be required. 
For this we assume a cm energy of 14 TeV. 

In practice, 
each experiment uses a specific decay channel for the $H$, as for
example, 
$H \to b\bar b$ \cite{CMSJHEP} or $H\to \gamma\gamma$ \cite{CMSPRL},
and the relevant
quantity is the cross section times the branching ratio for the decay
channel used.
We have concentrated on the $b\bar b$ decay
channel of the Higgs, as this has the largest branching ratio. But it
is obvious that more information can be obtained by combining it with
other cleaner channels like $H \to \gamma \gamma$ (with a somewhat low 
branching ratio, $2.27 \times 10^{-3}$). Thus, when we use the
term the cross
section below, we refer to the cross section for a $tb \bar b$ final state, or in case
of charged-lepton distributions, a $b \ell \nu b \bar b$ final state. We
combine cross sections for $tH$ and the conjugate $\bar tH$ 
production channels. We assume a Higgs mass of 125 GeV and a top mass of
172 GeV. We make use of  CTEQ\-6L parton distributions. Though other
parton distributions may predict somewhat different cross section
values, the main features of the kinematic distributions which we
examine
are not expected to change significantly.

\subsection{The total cross section}

As mentioned earlier, the cross section for $tH$ production does not depend
on the chirality of the couplings. Hence, for the cross section, we need only
the value of the relevant $\kappa_{Hqt}$. 
For $\kappa_{Hut}=1$ and $\kappa_{Hct}=0$, we obtain the cross sections for 
$tH+\bar t H$ production with $H$ decaying into
$b\bar b$ as 15.1 pb for the sum of the $u$ and $\bar u$ contributions.
For $\kappa_{Hut}=0$ and $\kappa_{Hct}=1$ we obtain 1.79 pb for the sum of 
the $c$ and $\bar c$ contributions. We assume
the $H\to b\bar b$ branching ratio to be 0.582. One could similarly
put in the relevant branching ratios in the case of other $H$ decays. 
Thus, if both $u$ and $c$ contributions are assumed to exist,
the experimental measurement of the cross section can only give a limit
on the combination 
\begin{equation}\label{sigmaHt}
\sigma_{Ht} = (15.1\,{\rm pb}) \,\kappa_{Hut}^2 + (1.79\,{\rm pb}) \, 
\kappa_{Hct}^2.
\end{equation}

\subsection{Top-quark polar distribution}
We next calculate the top polar-angle 
differential cross section contributions from the $u$ and $c$ quarks 
in the laboratory (lab) frame
in order to
investigate if the shape can be used to determine their relative
contributions.

\begin{figure}[t]
\centering
\includegraphics[width=0.8\columnwidth]{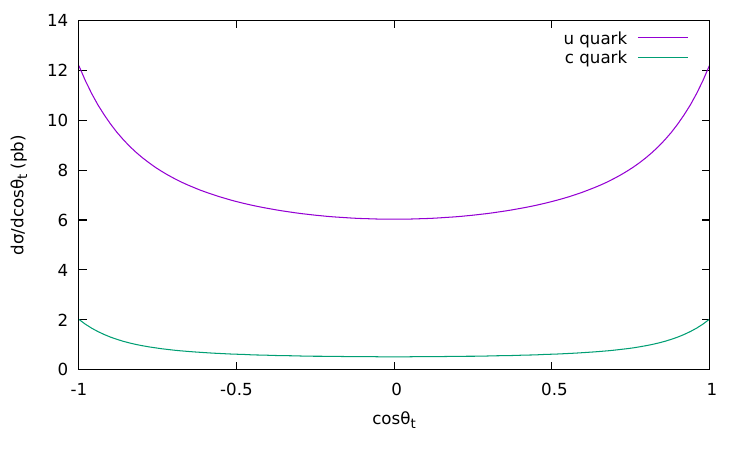}
\caption{Top-quark polar distributions for unit $Hqt$ coupling, when
$q$ is the $u$ quark, and when $q$ is the $c$ quark.}
\label{topcthdist}
\end{figure}

The lab frame 
differential cross section $d\sigma/d\cos\theta_t$ is plotted as a
function of $\cos\theta_t$ in Fig. \ref{topcthdist}, for the $u$-quark
contribution for $\kappa_{Hut} =1$ and for the $c$-quark contribution
for $\kappa_{Hct}=1$. 
\begin{figure}[t]
\centering
\includegraphics[width=0.8\columnwidth]{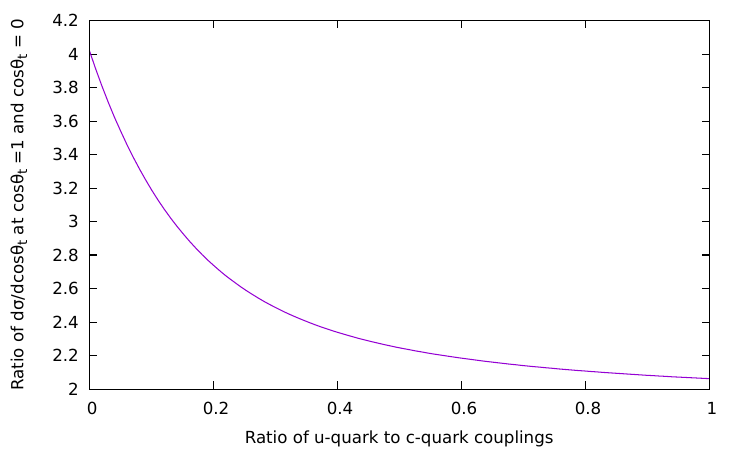}
\caption{The ratio of the value of $d\sigma/d\cos\theta_t$ at
$\cos\theta_t = 1$  to its value at $\cos\theta_t = 0$
plotted as a function of $\kappa_{Hut}^2/\kappa_{Hct}^2$
}
\label{curatio}
\end{figure}
We find that the $\cos\theta_t$ distribution is shallower for the 
$u$-quark contribution
as compared to the $c$-quark contribution. The differential
cross section at $\cos\theta_t = \pm 1$ is about 4 times that at 
$\cos\theta_t = 0$ for the $c$-quark contribution, 
whereas the ratio of differential cross sections is about 2 for the 
$u$-quark contribution.  
This is illustrated in Fig. \ref{curatio}, where this ratio is
plotted as a function of the coupling ratio
$\kappa_{Hut}^2/\kappa_{Hct}^2$.

By measuring the ratio of the differential cross sections at these
values of top polar angles, one can infer the relative contributions of
the the $u$-quark and $c$-quark couplings to $tH$. 
We take up a concrete example as an illustration.
If the ratio of differential cross sections at $\cos\theta_t =  1$ and 
$\cos\theta_t = 0$ is found to be 3, then Fig. \ref{curatio} shows that 
the ratio 
$\kappa_{Hut}^2/\kappa_{Hct}^2$ is about 0.15. 
If the experimental value is, let us say, 0.01 pb for the cross
section with a $b\bar b$ final state arising from $H$ decay, 
then, using this value in eq. (\ref{sigmaHt}),
the value of
$\kappa_{Hct}^2$ is about 0.01/4.04, or very nearly $2.5\times 10^{-3}$, and that of
$\kappa_{Hut}^2$ is $0.15\times (2.5\times 10^{-3})$, which is about 
$3.75\times 10^{-4}$.


It is difficult to make an estimate of the accuracy of such a
determination, since both coupling constants are unknown. Nevertheless,
to attempt a rough estimate in the example considered above, 
we can see that the assumed
cross section of 0.01 pb for a $Hb\bar b$ final state
would enable $10^4$
events to be produced for an integrated luminosity of 1000 fb$^{-1}$,
amounting to an uncertainty of 1\%. Taking this as the dominant error, the
uncertainty in the determination of the couplings would be about 0.5\% in a
typical case, with an integrated luminosity of 1000 fb$^{-1}$.

\subsection{Charged-lepton angular distributions}

We have worked out the differential cross section for the charged lepton
arising from $t$ decay in the lab frame. 
We choose a frame where the beam momentum
direction is the $z$ axis, and
the top momentum direction lies in the $xz$ plane in the laboratory
 frame.

\begin{figure}[t!]
\centering
\includegraphics[width=0.8\columnwidth]{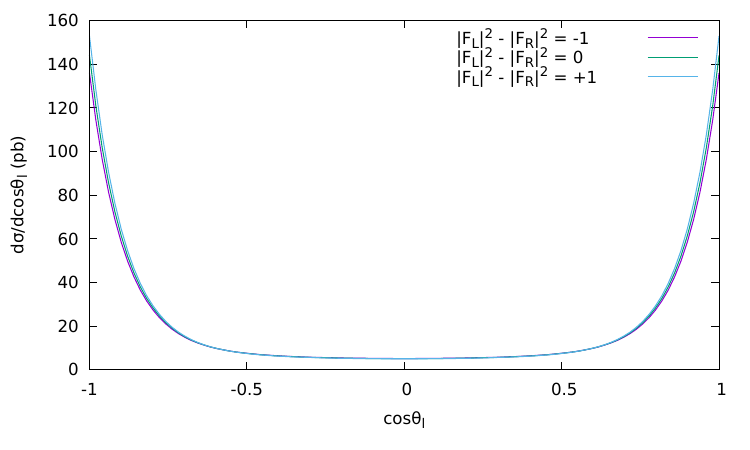}
\caption{The charged-lepton polar angle distribution in the $ug \to tH$
process for $\kappa_{Hut}=1$, and different combinations of left-handed
and right-handed coupling combinations.}
\label{ucthdist}
\end{figure}

We find that 
the $\cos\theta_{\ell}$ distribution is very nearly 
independent of the value
of $|F_L^{Hq}|^2 - |F_R^{Hq}|^2$ for the
$u$-quark as well as the $c$-quark contributions.
This is illustrated for the $u$-quark process in Fig. \ref{ucthdist}.
(We drop the superscripts $Hu$ and $Hc$ on $F_L$ and
$F_R$.)
We thus would not get much information on the chiral structure of the
couplings from the polar distribution.

\begin{figure}[h!]
\centering
\includegraphics[width=0.8\columnwidth]{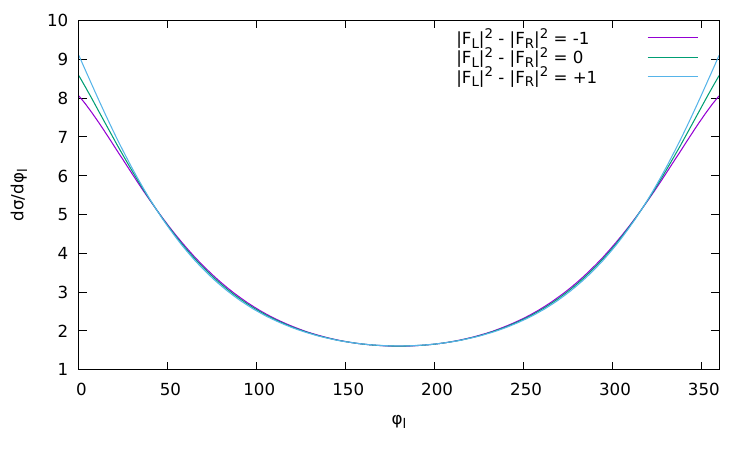}
\caption{The charged-lepton azimuthal distribution from the 
$ug \to tH$ subprocess contribution for
$\kappa_{Hut}=1$, and for 3 different combinations of left-handed
and right-handed coupling combinations.}
\label{uphidist}
\end{figure}

\begin{figure}[h!]
\centering
\includegraphics[width=0.8\columnwidth]{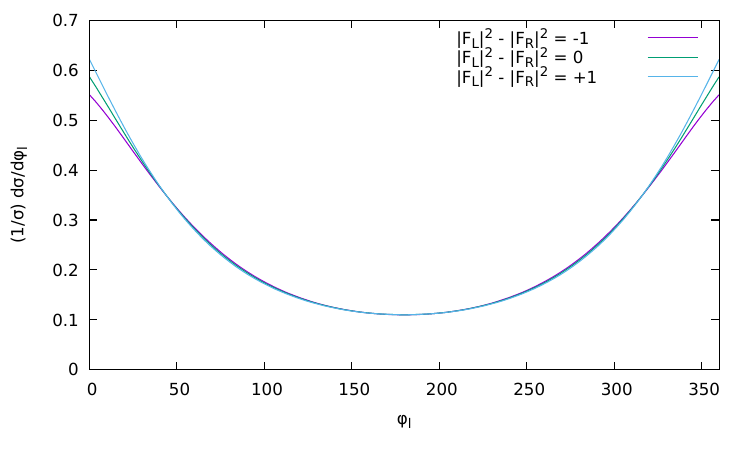}
\caption{The charged-lepton azimuthal distribution normalized to the
total cross section $\sigma$ in the $ug \to tH$ subprocess contribution 
for $\kappa_{Hut} = 1$, and for 3 different combinations of left-handed
and right-handed coupling combinations.}
\label{unormphidist}
\end{figure}

As for the azimuthal distributions, we do find some dependence on the
chiral structure of the couplings.
Fig. \ref{uphidist} shows the differential $\phi_{\ell}$ distribution
for the $u$-quark contribution as a function of $\phi_{\ell}$, whereas
Fig. \ref{unormphidist} shows the $\phi_{\ell}$
distribution, normalized to the total cross section $\sigma$.
For increasing $|F_L|^2 - |F_R|^2$,
the $\phi_{\ell}$ distribution for the $u$-quark contribution shows a
steady increase in the value at $\phi_{\ell} = 0, 2\pi$, while the
value at $\phi_{\ell} = \pi $ increases very slowly. The increase in
the value of the differential cross section at $\phi_{\ell} = 0, 
2\pi$ in going
from $|F_L|^2 - |F_R|^2 = -1$ to $|F_L|^2 - |F_R|^2 = 1$ 
is about 10\%, whereas the corresponding increase for
$\phi_{\ell} = \pi$ is about 1\%.
This also means that the ratio of the cross sections at $\phi_{\ell} = 0, 
2\pi$ and $\phi_{\ell} = \pi$ changes by about 10\% in going from 
$|F_L|^2 - |F_R|^2 = -1$ to $|F_L|^2 - |F_R|^2 = 1$.
For the $c$-quark contribution, the changes are negligible for
$\kappa_{Hct} = 1$.
This implies that if $\kappa_{Hut}$  is known to be much
larger than $\kappa_{Hct}$, measurement of the differential cross
section at $\phi_{\ell} \approx 0$ or $2\pi$ with 10\% accuracy or
better may be able to distinguish between right chiral and left chiral
couplings. In the presence of a comparable $\kappa_{Hct}$
contribution, with both $\kappa_{Hut}$ and $\kappa_{Hct}$ assumed small 
compared to 1, this
distinction will not be possible.  

To make a rough estimate of the luminosity that would be required for 
distinguishing the
chirality of the $\kappa_{Hut}$ couplings, we note that for a range of
$\Delta\phi_{\ell}$ around $\phi_{\ell}=\pi$, the number of events in 
such a
bin would be approximately $\Delta N \approx 8.5\,\Delta\phi_{\ell}\,
{\cal L}$, where ${\cal
L}$ is the luminosity in pb$^{-1}$. 
For a fractional change of 10\% in the
differential cross section due to the difference between left-handed and
right-handed chirality 
structures to be measurable at the 1 $\sigma$ level, 
$\Delta N$ should be
larger than 100. For a choice of $\Delta \phi_{\ell} = 0.1$ radian
(approximately 6$^{\circ}$), the required luminosity is about 118
pb$^{-1} \approx 0.1 {\rm fb}^{-1}$. 
This assumed $\kappa_{Hut}=1$. Assuming a limit on $\kappa_{Hut}$ of
about 0.01 with a future higher luminosity run (the current limit being 
of the order of 0.037), the required
luminosity would be 1000 fb$^{-1}$. 

We note in passing that the angular distributions of the charged lepton
arising in top decay can be used as a measure of the top
polarization in its production process. Of the polar and azimuthal
angular distributions, the azimuthal distribution is more sensitive to
the top polarization \cite{toppolazi}, though both of them are 
insensitive to any anomalous $tbW$ coupling at linear order in the 
decay process \cite{toppolazi,tbW}. Thus, the measurement of the 
angular distributions of charged leptons we have discussed here would 
not get contribution from anomalous $tbW$ couplings, if they are small.

\section{Conclusions}
We have studied the role that angular distributions of the top quark,
and those of the charged leptons from top-quark decay,
in the process of $tH$ production in $pp$ collisions, may play
in studying details of possible flavour-changing $Hqt$ ($q=u,c$)
couplings. 

Measurement of the cross section, carried out using  the LHC data at cm
energy of 13 TeV, enabled a limit to be put on the $Hqt$ coupling
$\kappa_{Hqt}$ assuming contribution from only one of the quarks $u$ or
$c$. We have studied the role that
the polar distribution of the top quark can play in determining the
relative contributions of the $Hut$ and $Hct$ couplings. In particular,
since the shapes of the polar distributions arising from $u$ and $c$
contributions are different, the ratio of differential cross sections 
at $\theta = 0$ and $\theta = \pi/2$, together with the total cross
section, can identify, given sufficient data, the ratio
$\kappa_{Hut}/\kappa_{Hct}$. The chirality of the coupling does not play
any role in this because the differential cross section does not depend
on it. A rough estimate of the uncertainty in a typical choice of couplings
yields 0.5\% for an integrated luminosity of 1000 fb$^{-1}$.

We have then studied the possibility of the angular distribution of the 
charged lepton helping to unravel the chiral structure of the $Hqt$
coupling. We find that polar distributions of the charged leptons 
are insensitive to the chiral structure. Of the azimuthal distributions
arising from the $u$ contribution and the $c$ contribution, the
distribution from the $u$ contribution is somewhat sensitive to the
chiral structure. In particular, the differential cross section near
$\phi_{\ell} = 0$ or $\pi$ varies by about 10\% in going from
a left chiral coupling to a right chiral coupling, whereas the differential
cross section near $\phi_{\ell}=\pi/2$ is almost unchanged.  
A rough estimate of the integrated luminosity which may be needed to
distinguish this 10\% variation yields 1000 fb$^{-1}$. 

These results assume certain ideal experimental conditions, and one
needs to exercise caution in accepting them as such for the following
reasons. 

The measurement of the top angular distribution relies on accurate
reconstruction of the top four-momentum, 
which presents a challenge at LHC experiments. 
In case of leptonic decays of the top, the unobserved
neutrinos complicate the reconstruction of the top four-momentum.
The  hadronic top decays suffer from QCD and combinatorial backgrounds,
even though the hadronic decays give better statistics because of the 
larger branching ratio. Recent results in machine
learning-based top tagging have demonstrated significant improvements in
top tagging performance over the standard jet substructure methods used 
in past. These would indeed play a useful role in improving 
the measurement of top
angular distributions suggested in this work. A couple of recent
relevant papers 
on the use of machine-learning for top tagging are
\cite{sahu,
Keicher:2023mer}
which are also good sources of references to earlier work
on the topic.

Another point to be borne in mind is that the 
geometry of the LHC detectors imposes limitations on the detection of 
particles in
the forward direction. It may therefore be unrealistic to expect the 
detection of top-quark events with $\cos\theta_t = \pm 1$. Clearly, one would have to work out the results for more
realistic forward and backward values of the top polar angle.  

We hope that a more detailed numerical analysis with realistic
experimental conditions will enable use of the above observations for
measuring (or limiting) more details of the structure of the flavour
changing Higgs couplings at future LHC runs with higher luminosity.

\section*{Acknowledgement}
The author gratefully acknowledges financial support from the Senior 
Scientist programme of the Indian National Science Academy, New Delhi.
The author thanks the Indian Academy of Sciences, Bengaluru, for
hospitality and a pleasant working environment during the completion of
this work.



\end{document}